\documentclass[twocolumn]{aastex63}

\pdfoutput=1
\usepackage{subcaption}

\received{2023 July 4}
\revised{2023 October 20}
\accepted{2023 November 7}

\submitjournal{ApJ}

\shorttitle{Pulsation in TESS Objects of Interest}
\shortauthors{Gomes et al.}

\begin{document}

\title{Pulsation in TESS Objects of Interest}

\correspondingauthor{José Renan de Medeiros}
\email{renan@fisica.ufrn.br}
\author[0000-0002-2023-7641]{R. L. Gomes}
\affiliation{Departamento de F\'isica Te\'orica e Experimental, Universidade Federal do Rio Grande do Norte, Campus Universit\'ario, Natal, RN, 59072-970, Brazil}
\author[0000-0001-5578-7400]{B. L. Canto Martins}
\affiliation{Departamento de F\'isica Te\'orica e Experimental, Universidade Federal do Rio Grande do Norte, Campus Universit\'ario, Natal, RN, 59072-970, Brazil}
\author[0000-0002-3916-6441]{D. O. Fontinele}
\affiliation{Departamento de F\'isica Te\'orica e Experimental, Universidade Federal do Rio Grande do Norte, Campus Universit\'ario, Natal, RN, 59072-970, Brazil}
\author[0000-0001-5845-947X]{L. A. Almeida}
\affiliation{Escola de Ci\^encia e Tecnologia, Universidade Federal do Rio Grande do Norte, Campus Universit\'ario, Natal, RN, 59072-970, Brazil}
\author[0000-0001-5369-1085]{R. Alves Freire}
\affiliation{Departamento de F\'isica Te\'orica e Experimental, Universidade Federal do Rio Grande do Norte, Campus Universit\'ario, Natal, RN, 59072-970, Brazil}
\author[0000-0003-2719-8056]{A. C. Brito}
\affiliation{Instituto Federal do Cear\'a (IFCE), Campus Sobral, Av. Dr. Guarani, 317 - Derby Clube, Sobral, CE, 62042-030, Brazil}
\author[0000-0002-6085-5962]{R. G. S. B. de Amorim}
\affiliation{Departamento de F\'isica Te\'orica e Experimental, Universidade Federal do Rio Grande do Norte, Campus Universit\'ario, Natal, RN, 59072-970, Brazil}
\author[0000-0002-8525-7977]{C. E. Ferreira Lopes}
\affiliation{Instituto de Astronom\'ia y Ciencias Planetarias, Universidad de Atacama, Copayapu 485, Copiap\'o, Chile}
\affiliation{Millennium Institute of Astrophysics, Nuncio Monse\~nor Sotero Sanz 100, Of. 104, 7500000 Providencia, Santiago, Chile}
\author[0000-0001-8218-1586]{D. Hazarika}
\affiliation{Instituto de Astronom\'ia y Ciencias Planetarias, Universidad de Atacama, Copayapu 485, Copiap\'o, Chile}
\affiliation{Millennium Institute of Astrophysics, Nuncio Monse\~nor Sotero Sanz 100, Of. 104, 7500000 Providencia, Santiago, Chile}
\author[0000-0001-8218-1586]{E. Janot-Pacheco}
\affiliation{Departamento de Geof\'isica Te\'orica e Ci\^encias Atmosf\'ericas, Universidade de S\~ao Paulo, Rua do Mat\~a, Campus Universit\'ario, 1226, 05508-090, São Paulo, SP, Brazil}
\author[0000-0001-5845-947X]{I. C. Le\~ao}
\affiliation{Departamento de F\'isica Te\'orica e Experimental, Universidade Federal do Rio Grande do Norte, Campus Universit\'ario, Natal, RN, 59072-970, Brazil}
\author[0000-0002-2425-801X]{Y. S. Messias}
\affiliation{Departamento de F\'isica Te\'orica e Experimental, Universidade Federal do Rio Grande do Norte, Campus Universit\'ario, Natal, RN, 59072-970, Brazil}
\author[0000-0002-5955-5882]{R. A. A. Souza}
\affiliation{Departamento de F\'isica Te\'orica e Experimental, Universidade Federal do Rio Grande do Norte, Campus Universit\'ario, Natal, RN, 59072-970, Brazil}
\author[0000-0001-8218-1586]{J. R. De Medeiros}
\affiliation{Departamento de F\'isica Te\'orica e Experimental, Universidade Federal do Rio Grande do Norte, Campus Universit\'ario, Natal, RN, 59072-970, Brazil}

\begin{abstract}

We report the discovery of three Transiting Exoplanet Survey Satellite Objects of Interest (TOI) with signatures of pulsation, observed in more than one sector. Our main goal is to explore how large is the variety of classical pulsators such as $\delta$ Sct, $\gamma$ Dor, RR Lyrae and Cepheid among TOI pulsators. The analysis reveals two stars with signatures of $\delta$ Sct and one of $\gamma$ Dor, out of a sample of 3901 TOIs with available light curves (LCs). To date, there is a very scarce number of known pulsating stars hosting planets. The present finding also emerges as an exciting laboratory for studying different astrophysical phenomena, including the effects of star-planet interaction on pulsation and timing detection of planetary companions. We have also identified 16 TOI stars with periodicities and LCs morphology compatible with different classical pulsating classes, but for most of them, the dominant frequency signals originate from contaminating sources.

\end{abstract}

\keywords{Transit photometry; Substellar companion stars; Star-planet interactions; Stellar pulsations}

\section{Introduction} \label{sec:intro}

Stellar pulsation represents a unique laboratory to test a variety of crucial questions concerning star evolution, including the distribution of material between core and envelope, stellar size determination, radial surface velocities,  mass-loss dynamics in AGB variables, and mass-transfer in close binary systems (e.g., \citealp{Chaplin2013, Aerts2021}). Pulsation and stellar rotation can act on the atmospheres to create other mixing effects, affecting atmospheric observables \citep{Anderson2016}. Stellar pulsation can also affect the efficiency of nucleosynthesis products measured in the atmosphere or during mass-loss (e.g., \citealp{Karakas2012,Stancliffe2013}). Stars with pulsation also represent interesting laboratories for planet detection, using light travel time variations, a phenomenon resulting from mutual gravitational interaction between a pulsating star and an orbital companion (\citealp{Wolszczan1992,Sigurdsson2003,Suleymanova2014,Starovoit2017, Hey2020}), or from the identification of orbital modulation of planets in very tight orbits, caused by the reflected light from the illuminated side of the planetary companion (e.g.: \citealp{Charpinet2011, Silvotti2014}).

A study by \citet{Essen2020} has identified nonradial pulsations in TIC 129979528, a star hosting a known planet ultra-hot Jupiter (WASP-33b), with periods comparable to the period of the primary transit, from observations collected by the Transiting Exoplanet Survey Satellite (TESS). Indeed, WASP-33b was the first transiting planet detected around a $\delta$ Sct star, representing a benchmark in the study of exoplanets \citep{Herrero2011}. In addition, \citet{Hey2021} reported 13 Kepler Objects of Interest that show both planetary transits and $\delta$ Sct pulsation signatures. The discovery of hybrid pulsators ($\gamma$ Dor - $\delta$ Sct) with planetary companions \citep{Bognar2015, Lampens2021}, namely stars in which both low radial order $p$- and high-order $g$-modes are self-excited at the same time, also represents new challenges for the theory of stellar pulsations. More recently, \citet{Kalman2023} reported, from TESS observations, the discovery of a substellar companion orbiting HD 31221, also found evidence that this star is a $\gamma$ Dor - $\delta$ Sct hybrid ($\gamma$ Dor - $\delta$ Sct) pulsator.

The TESS space mission \citep{Ricker2015} produces a photometric differential time series for hundreds of thousands of stars, primarily dedicated to searching for terrestrial planets transiting nearby bright stars. The large number of observed targets coupled with the high quality of the acquired data is opening new horizons for studying various astrophysical phenomena. Complex rotational modulation among M dwarfs \citep{Zhan2019}, rotation and pulsation of magnetic chemically peculiar A-type stars \citep{Cunha2019}, identification of flares in M-type stars \citep{Gunther2019,Doyle2020}, frequency variability of hot young $\delta$ Sct stars \citep{Antoci2019}, pulsation in main-sequence B stars and $\delta$ Sct, $\gamma$ Dor, and roAp candidates \citep{Balona2020}, rotation of B stars \citep{Barraza2022}, asteroseismology analysis of pulsating subdwarf B stars (e.g., \citealp{Sahoo2023}), characterization of exoplanets and their host stars using asteroseismology (e.g., \citealp{Huber2019}) and the detection of low-amplitude features like weak modulation, period jitter, and timing variations \citep{Plachy2021}, are among the most important TESS findings to date. Of course, these results are in parallel with major findings in exoplanetology, including transit detection of known exoplanets, the discovery of new exoplanets, identification of phase signatures and secondary eclipses, refinement of transit ephemeris, and asteroseismology as a pathway to improve stellar and planetary parameters \citep{Kane2021}.

The study by \citet{Canto2020}, dedicated to the determination of rotation period for TESS Objects of Interest (TOIs), has revealed the signature of potential pulsation among ten targets out of a sample of 1000 TOIs\footnote{\url{https://filtergraph.com/tess rotation tois}}. Aiming to increase the present-day sample of classical pulsating stars with confirmed planets or planet candidates, we present the classification of the pulsation behavior of a unique sample of 24 TOI stars, including eight pulsator candidates from \citet{Canto2020} and 16 TOI pulsator candidates identified in this study. Specifically, our study consists of searching for A-F type classical pulsators, such as $\delta$ Sct, $\gamma$ Dor, RR Lyrae, and Cepheids, located in the lower part of the instability strip at different evolutionary stages, from the main-sequence to the giant branch \citep[e.g.,][]{Papics2013, Aerts2021, Kirmizitas2022}. We organized the paper as follows. Section \ref{observation} presents the stellar sample and observational data set used in this study. Section \ref{sec:results} provides the main results. We present a summary in Section \ref{sec:summary}.

\section{Stellar Sample and Observational Data} \label{observation}

For the present purpose, we have analyzed a sample of 2901 TOIs in the search for pulsation signatures from their light curves (LCs), acquired on 2 minutes cadence mode. We downloaded these data from the {\em FFI-TP-LC-DV Bulk Downloads Page} of the Mikulski Archive for Space Telescopes\footnote{\url{https://archive.stsci.edu/tess/bulk_downloads.html}} using the cURL scripts available for retrieving PDCSAP reduced LCs. \citet{Jenkins2016} described the TESS Science Processing Operations Centre (SPOC) pipeline that produces the 2 minute LCs. We performed additional processing on these LCs, when required, to avoid possible distortions in the signature of periodicities resulting from outlier removal and instrumental trend, following the recipe by \citet{deMedeiros2013}, \citet{PazChinchon2015} and \citet{Canto2020}, plus removal of transits following the procedure described in \citet{PazChinchon2015}.

\citet{Canto2020} used the {\em MAPS} package in the analysis of the post-processed LCs, which is a manifold interactive platform built in the {\em Interactive Data Language} (IDL\footnote{\url{https://www.l3harrisgeospatial.com/Software-Technology/IDL}}; \citealp{Landsman1995}) that contains several tools for post-treatment and analysis of time series, including Lomb-Scargle periodograms (e.g., \citealp{Scargle1982,Horne1986,Press1989}), Fast Fourier Transform (FFT) (e.g., \citealp{Zhan2019}) and wavelet analysis (e.g., \citealp{Grossmann1984,Bravo2014}). The he wavelet analysis identifies periodicity associated with the persistence of the phenomenon along the observed time range. Using the {\em MAPS} package, we computed the Lomb-Scargle and FFT frequency spectra, as well as wavelet maps of each TOI LC, from which we extracted the frequencies and amplitudes of the strongest peaks above 4~day$^{-1}$. False alarm probabilities (FAPs) were computed for periodogram peaks based on the Lomb-Scargle method, using Equation (22) of \citet{Horne1986}. Only peaks with FAP less than 1\%, corresponding to significance levels greater than 99\%, were considered. The errors on the peak frequencies were estimated using Equation (2) of \citet{Lamm2004}.

To identify the TOI pulsator candidates, we followed the same procedure applied by \citet{Canto2020}, separating the stars exhibiting LCs with unambiguous photometric variability into two groups:  (i) those with a typical rotational modulation in the LC, namely a signature characterized by semi-regular flux variability that use to be multi-sinusoidal, most commonly showing single or double dips per rotation cycle (e.g., \citealp{Lanza2003,Lanza2007,deMedeiros2013,Basri2018,BasriNguyen2018}), and (ii) those with a typical pulsation profile, namely those stars with LCs displaying a more regular shape of the flux variation, sometimes with constant amplitude, or a regular amplitude variation usually forming steady beats (e.g., \citealp{FerreiraLopes2015}). However, some pulsators, such as $\gamma$ Dor variables, may present irregularities in their LCs that can be confused with rotational modulation, and an additional control should be applied, typically the identification of an asymmetry in their variability signatures skewed to higher fluxes. This analysis revealed 16 new pulsator candidates out of the sample of 2901 TOIs.

We have merged this list of 16 new pulsator candidates with eight pulsator candidates from \citet{Canto2020}, composing a sample of 24 TOI stars with likely pulsation signatures, which are listed in Table \ref{table1} with their respective parameters. The referred stars were subjected to a detailed check to verify if the pulsation signals obtained in our study were genuinely from the considered TOI. Indeed, given the TESS large plate scale of 21$''$ pixel$^{-1}$, analysis of signals obtained from the LCs of this mission should be made with caution due to potential contamination from blended LCs, which can induce an interpretation of the observed photometric variability to a wrong source (\citealp{Mullally2022,Higgins2023,Pedersen2023}).
As a first step to assess the potential impact of contamination by nearby sources, we analyzed the crowding metric (CROWD) calculated by the SPOC pipeline (\citealp{Caldwell2020b}), which indicates the fraction of the light in the TESS aperture that comes from the target star given the positions and amplitudes of stars in the TESS Input Catalog. A crowd value near 1.0 indicates a potentially isolated star, while lower values indicate significant crowding from neighbors. This analysis found five stars with CROWD $<$ 0.9 within the total sample of 24 stars. (Note that the CROWD values shown in Table \ref{table1} are an average of this parameter for stars observed in more than one sector.)

\begin{table*}[!htb]
\begin{center}
\movetableright=-1in
	\caption{TOI Stars with Pulsation Signature}
\label{table1}
\begin{tabular}{cccccccc}
\hline \hline
TIC & $T_{eff}$ & $\log g$ & Lum & $P_{orb}$ & CROWD & TESS Sectors & References \\
& (K) & ($cm/s^2$) & ($L_{\sun}$) & (days) &   &   &   \\
\hline
58533991	&	7423	&	3.898	&	16.136	&	3.209	&	0.968	&	37	&	1,2	\\
96246348	&	11018	&	4.2196	&	57.830	&	1.409	&	0.754	&	33, 34, 61	&	2,3	\\
97700520	&	15666	&	3.7992	&	965.900	&	5.426	&	0.995	&	33, 34	&	2,3	\\
103195323	&	7437	&	4.088	&	10.481	&	5.590	&	0.975	&	25, 26, 52, 53, 59	&	1,2	\\
107782586	&	15389	&	3.927	&	421.400	&	1.960	&	0.978	&	34, 61	&	2,3	\\
118084044	&	7703	&	3.500	&	49.771	&	2.744	&	0.990	&	35-37, 61, 62	&	1,2	\\
123898871	&	7245	&	4.191	&	7.151	&	4.909	&	0.970	&	33	&	1,2	\\
129881395	&	5982	&	4.172	&	2.341	&	3.363	&	0.188	&	61	&	1,2	\\
144043410	&	6888	&	3.993	&	8.476	&	6.264	&	0.794	&	34	&	1,2	\\
149833117	&	6578	&	4.310	&	3.099	&	4.052	&	0.917	&	20, 47, 60	&	1,2	\\
150299840	&	6832	&	4.021	&	7.596	&	19.474	&	0.946	&	28-36, 38, 39, 61, 62	&	1,2	\\
156987351	&	7451	&	4.047	&	11.697	&	3.063	&	0.997	&	6, 7, 33, 34, 61	&	1,2	\\
164173105	&	7164	&	4.257	&	5.763	&	3.073	&	0.991	&	16, 56	&	1,2	\\
171160243	&	7825	&	4.072	&	14.586	&	1.389	&	0.940	&	43,44, 59	&	1,2	\\
179580045	&	7546	&	4.243	&	8.021	&	62.163	&	0.915	&	61, 62	&	1,2	\\
201604954	&	5600	&	4.340	&	1.099	&	4.606	&	0.867	&	13	&	1,2	\\
281716779	&	13452	&	3.901	&	468.500	&	8.170	&	0.966	&	33	&	2,3	\\
287196418	&	6982	&	4.197	&	5.718	&	3.695	&	0.990	&	14, 16, 17, 21, 26, 40, 41, 47, 49-51, 53-60	&	1,2	\\
297967252	&	12729	&	3.947	&	217.900	&	9.683	&	0.995	&	9, 10, 35, 36, 62	&	 2,3	\\
329277372	&	5780	&	4.438	&	-	&	2.888	&	0.718	&	16, 17, 56, 57	&	2	\\
333607525	&	6600	&	4.170	&	-	&	3.202	&	0.907	&	33	&	2	\\
374095457	&	9999	&	3.278	&	250.800	&	0.784	&	0.952	&	9, 10, 36, 37	&	2,3	\\
436873727	&	6798	&	4.114	&	5.924	&	22.097	&	0.998	&	18, 42-44, 58	&	1,2	\\
468997317	&	12705	&	3.777	&	280.000	&	7.503	&	0.920	&	9, 10, 35, 36, 62	&	2,3	\\

\hline \hline

\end{tabular}
\end{center}
\footnotesize{\textbf{Note}: The following information is listed: the TIC ID, effective temperature ($T_{eff}$), surface gravity ($\log g$), orbital period ($P_{orb}$) and TESS Sectors of observations. Values for $\log g$, luminosity (Lum) and $P_{orb}$ are rounded to three decimal digits. References for $T_{eff}$, $\log g$, Lum and $P_{orb}$: (1) \cite{Stassun2019}, (2) TOI catalog: \url{https://tess.mit.edu/toi-releases/}, (3) \cite{Gaia2023}. }
\end{table*}

After using the CROWD tool, we performed an additional check for potential contamination using \texttt{TESS\_Localize}, which is a powerful tool for pinpointing the true source of an observed variability \citep{Higgins2023}. Essentially, the \texttt{TESS\_Localize} method utilizes TESS pixel response function models to characterize systematics in the residuals of fitting the models to data. This method shows that even stars of more than three pixels outside a photometric aperture can produce significant contaminant signals in the extracted LCs.
The \texttt{TESS\_Localize} procedure takes advantage of the fact that a variable source contributes to the flux distribution in each pixel of a detector proportionally to the variability amplitude. As such, a set of input frequencies (in our case, the pulsating frequencies detected from a reduced LC) is used to simulate how a source variability affects the flux distribution across the detector pixels. The flux-distribution model is then fitted to the Target Pixel File (TPF) to precisely locate the source of variability, accurate to less than one-fifth of a pixel. The refined source location is cross-referenced with Gaia nearby sources to confirm whether the variability originates from the target or from a neighboring object.

An important output of the \texttt{TESS\_Localize} procedure is the ``relative likelihood'' (RL) parameter as an indicator of a certain target being the true source of variability. Based on the discussion presented by \citet{Higgins2023}, we consider the variability to originate from the target if RL is greater than 90\%. Otherwise, we assume that the variability may originate from another source and, in turn, the target to be suspicious of contamination. To refine our analysis, we also performed a visual inspection of the TPFs generated by \texttt{TESS\_Localize} to investigate the neighborhood of the target under study and to verify whether the target location visually matches the location of the variable source.
After this step, five stars were identified as potentially free of contamination, namely TIC 58533991, TIC 103195323, TIC 118084044, TIC 156987351, and TIC 164173105.

For our final sample selection, we also verified the possibility of the TOI status being false positives. Essentially, we performed a double-check of the apparent transit signals in the TESS LCs, combined with additional information from the literature, to identify any possibility of those TOIs being eclipsing binaries themselves or contaminated by eclipsing binaries. Based on this analysis, two objects were excluded, TIC 58533991 and TIC 156987351, which are described in more details  in Sect.~\ref{subdubious}. Our final sample is then composed of the three remaining stars potentially uncontaminated in both pulsating and transit signals: TIC 103195323, TIC 118084044, and TIC 164173105.

Table \ref{table2} lists the three bonafide TOI pulsator candidates, namely those stars with neither contamination nor offset effects, with their more significant extracted frequencies. Figure \ref{mapsbonafide} displays the TESS LC, corresponding phase-folded LC, FFT, Lomb–Scargle frequency spectra, and wavelet map results of the present analysis for the TIC 103195323, TIC 118084044, and TIC 164173105, respectively, although Figure 2 displays the heat maps from \texttt{TESS\_Localize} analysis.

\begin{table*}[!htb]
\begin{center}
\centering
\movetableright=-1.5in
\caption{The List of Extracted more Significant Frequencies for the Three Bonafide TOI Pulsator Candidates.}
\label{table2}
\resizebox{\textwidth}{!}{
\begin{tabular}{cccccccccccccccccc}
\hline \hline
TIC	&	$f_a $	&	$A_a$	&	$f_b $	&	$A_b$	&	$f_c $	&	$A_c$	&	$f_d $	&	$A_d$	&	$f_e $	&	$A_e$	&	$f_{orb}$	&	$f_a /f_{orb}$	&	$f_b /f_{orb}$	&	$f_c /f_{orb}$	&	$f_d /f_{orb}$	&	$f_e /f_{orb}$	&	Class	\\
	&	$\mu$Hz	&	$10^{-3}$	&	$\mu$Hz	&	$10^{-3}$	&	$\mu$Hz	&	$10^{-3}$	&	$\mu$Hz	&	$10^{-3}$	&	$\mu$Hz	&	$10^{-3}$	&	$\mu$Hz	&		&		&		&		&		&		\\ \hline
103195323	&	337.882	&	1.885	&	376.838	&	1.180	&	356.821	&	0.973	&	321.074	&	0.732	&	1.955	&	0.613	&	2.071	&	163.183	&	181.996	&	172.329	&	155.065	&	0.944	&	$\delta$ Sct 	\\
118084044	&	159.465	&	0.968	&	145.112	&	0.647	&	8478.216	&	0.564	&	4.316	&	0.355	&	158.806	&	0.275	&	4.217	&	37.813	&	34.409	&	2010.366	&	1.023	&	37.656	&	RR Lyrae	\\
164173105	&	19.731	&	15.142	&	19.023	&	5.744	&	20.359	&	3.791	&	18.058	&	3.614	&	10.949	&	3.392	&	3.767	&	5.238	&	5.050	&	5.405	&	4.794	&	2.907	&	 $\gamma$ Dor	\\
\hline

\end{tabular}}
\end{center}
\footnotesize{ \textbf{Note}: With one row for each TOI, the following information is listed: the TIC ID, frequencies $f_i$ and the corresponding amplitude $A_i$, orbital frequency $f_{orb}$, $f_{puls}/f_{orb}$ ratio and potential pulsation class.}
\end{table*}

\begin{figure*}[!htb]
	\centering
\begin{subfigure}[b]{0.74\textwidth}
     \includegraphics[width=\textwidth]{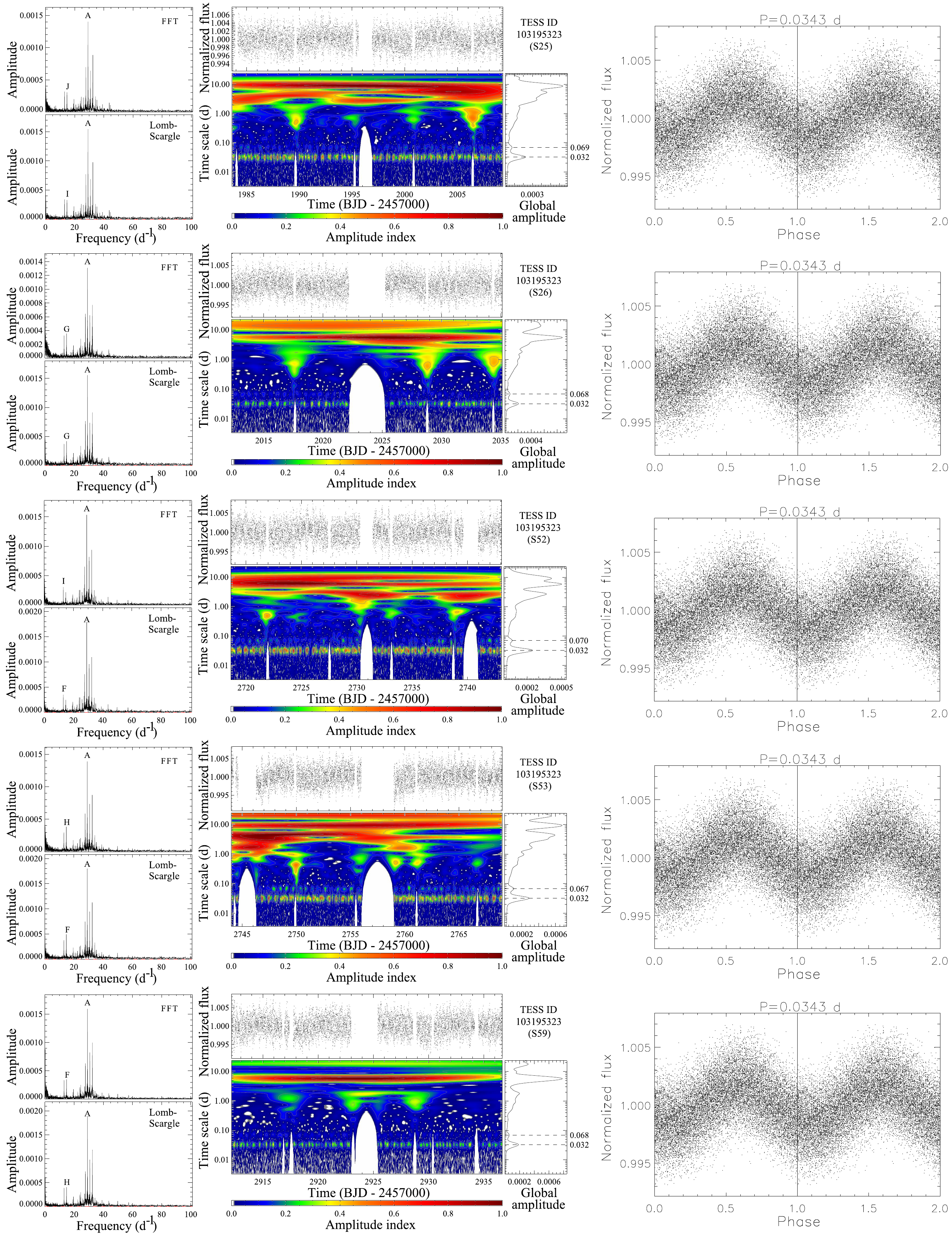}
     \subcaption{TIC 103195323}
     \label{map1031}
 \end{subfigure}
\caption{Analysis of the LC for: (a) TIC 103195323 (sectors 25, 26, 52, 53, and 59), (b) TIC 118084044 (sectors 35, 36, 37, and 61) and (c) TIC 164173105 (sectors 16 and 56). From left-hand to right-hand, FFT and Lomb-Scargle periodograms, TESS LC, wavelet maps, and the phase-folded LC.}
\end{figure*}

\begin{figure*}[!htb] \ContinuedFloat
    \centering
	\begin{subfigure}[b]{0.74\textwidth}
     \includegraphics[width=\textwidth]{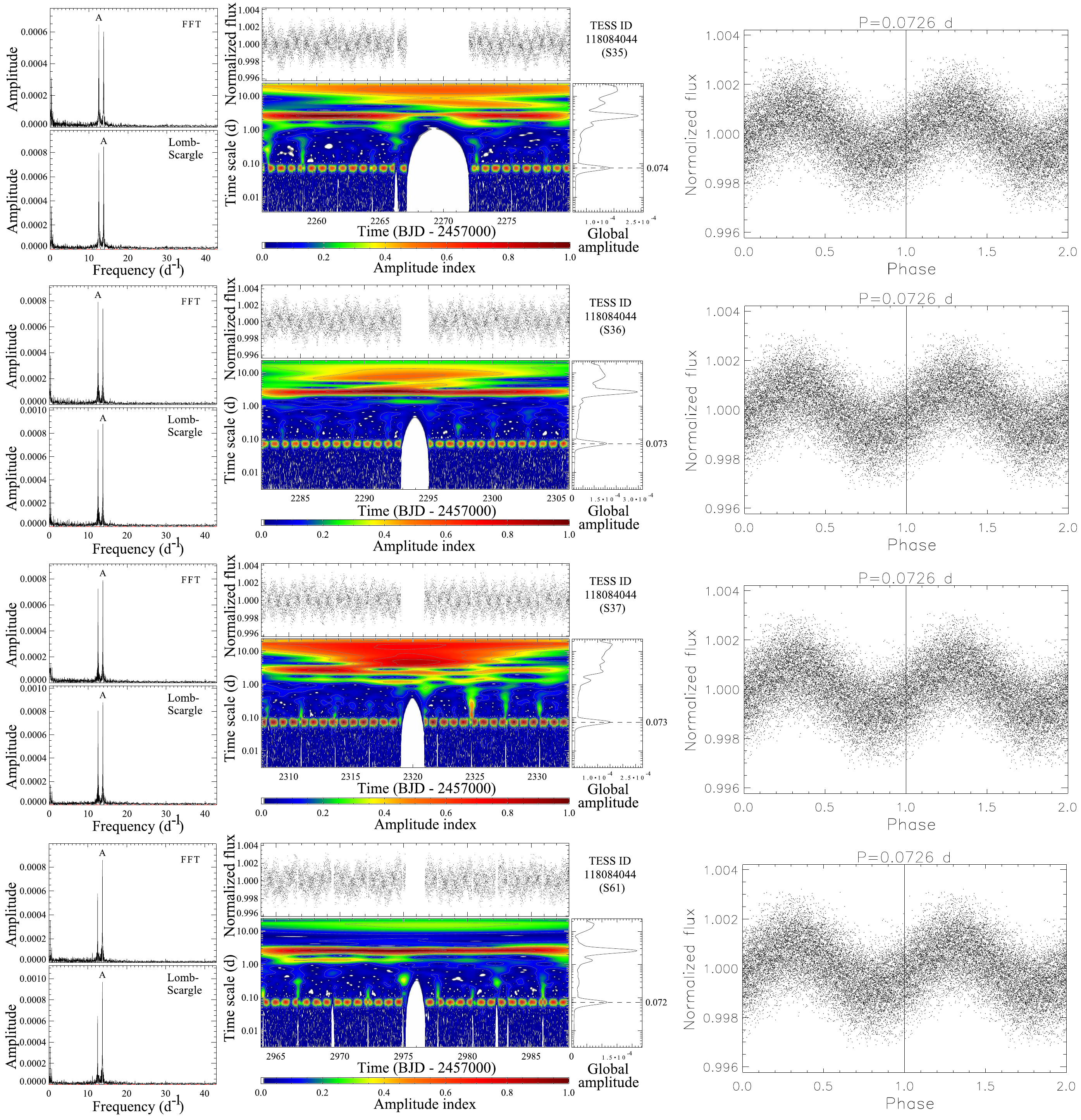}
     \subcaption{TIC 118084044}
     \label{map1180}
 \end{subfigure}
\end{figure*} 

\begin{figure*}[!htb] \ContinuedFloat
    \centering
	\begin{subfigure}[b]{0.74\textwidth}
     \includegraphics[width=\textwidth]{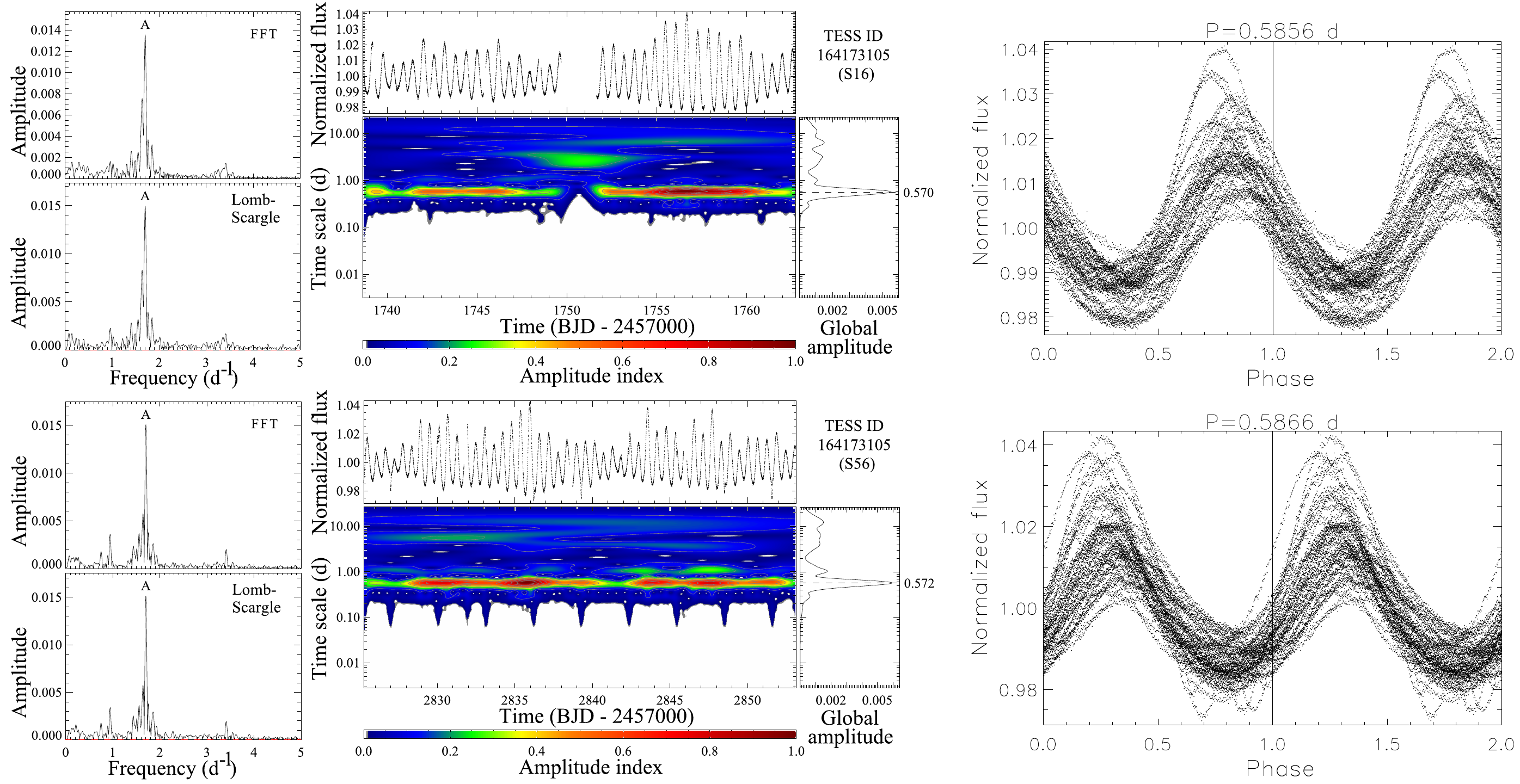}
     \subcaption{TIC 164173105}
     \label{map1641}
 \end{subfigure}

\caption{(Continued.)}
\label{mapsbonafide}
\end{figure*}

\begin{figure*}[!htb]
	\centering
 
    \begin{subfigure}[b]{1\textwidth}
     \includegraphics[width=\textwidth]{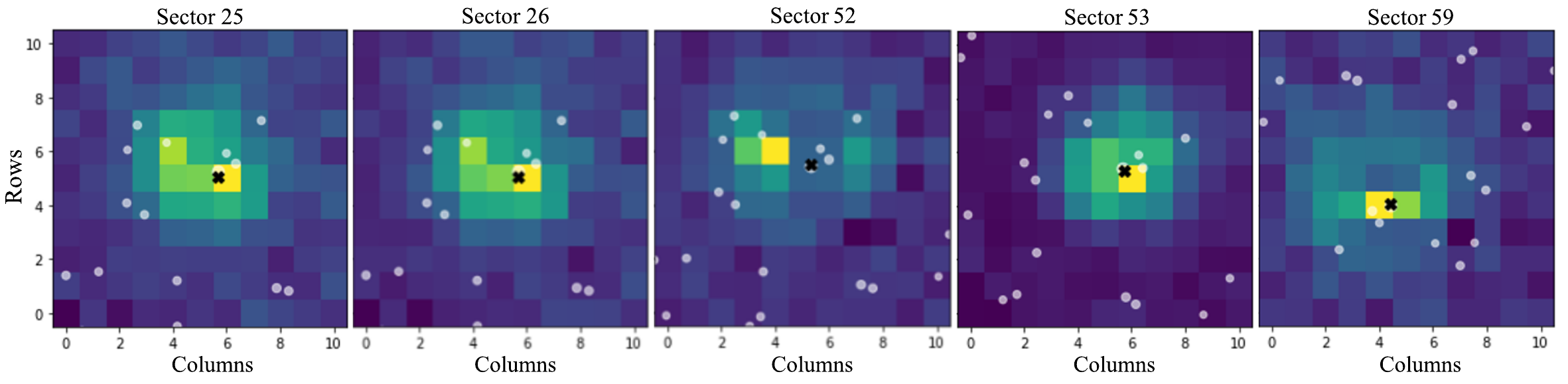}
     \subcaption{TIC 103195323}
     \label{tpf1031}
    \end{subfigure}

	\begin{subfigure}[b]{1\textwidth}
     \includegraphics[width=\textwidth]{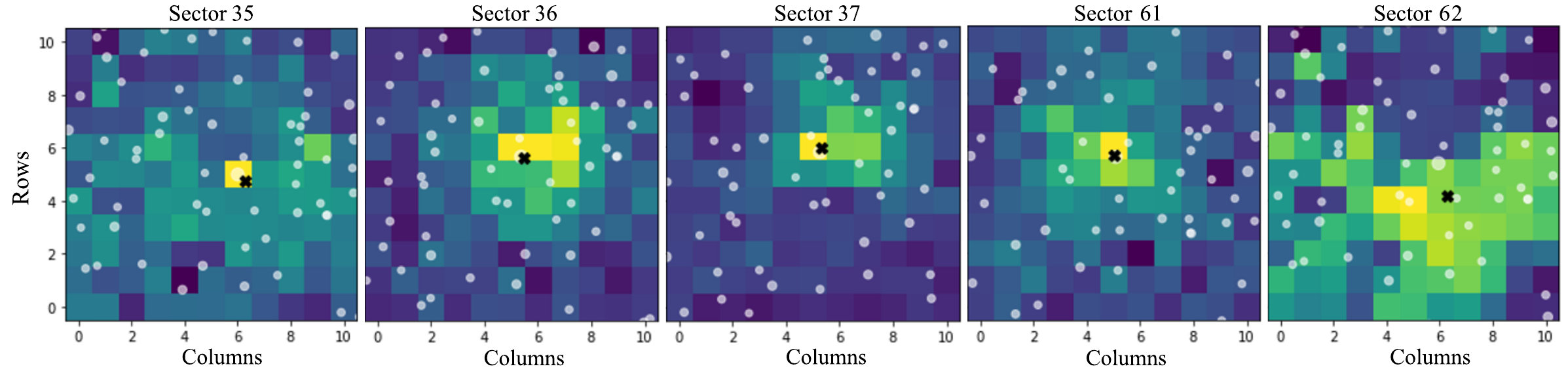}
     \subcaption{TIC 118084044}
     \label{tpf1180}
    \end{subfigure}

	\begin{subfigure}[b]{0.43\textwidth}
     \includegraphics[width=\textwidth]{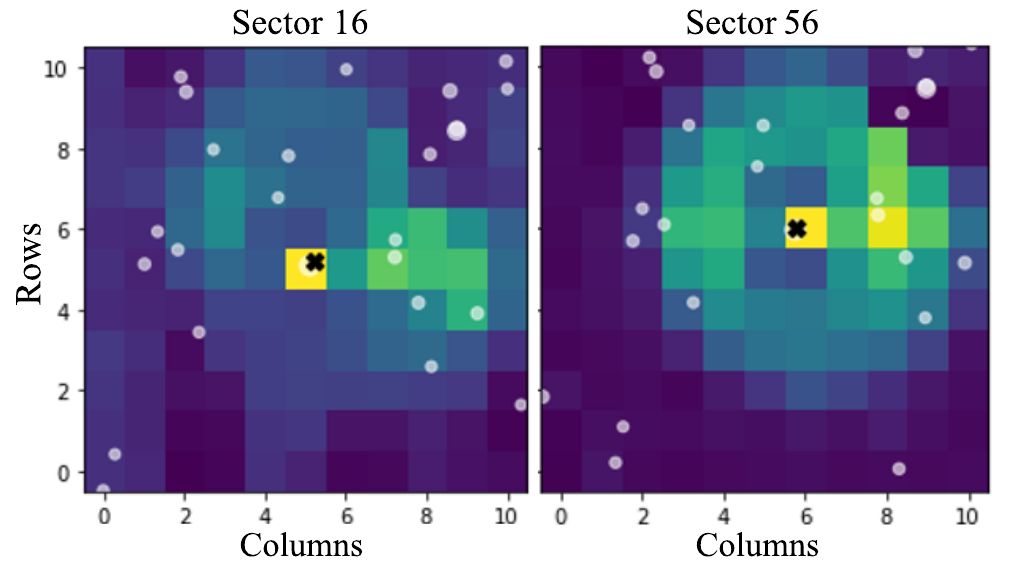}
     \subcaption{TIC 164173105}
     \label{tpf1641}
    \end{subfigure}

\caption{Heat maps generated by \texttt{TESS\_Localize} (\citealp{Higgins2023}) of the fitted amplitudes for each 21 pixel at the observed periods across the pixels downloaded from TESS for: (a) TIC 103195323 (sectors 25, 26, 52, 53, and 59), (b) TIC 118084044 (sectors 35, 36, 37, and 61) and (c) TIC 164173105 (sectors 16 and 56). The gray circles represent known Gaia Data Release 2 sources with T$_{mag}$ $>$ 15, approximately 3 mag dimmer than the target star. The black cross represents the best fit between the heat map and the TESS pixel response function (TESS PRF), as defined by \citet{Higgins2023}. In both cases, the best-fit location overlaps the Gaia location of our target at the center of the TESS pixels, indicating that the variability comes from the targeted source.}
\label{tpfbonafide}
\end{figure*}

\begin{figure*}[bt!]
	\centering
	\includegraphics[scale=.7]{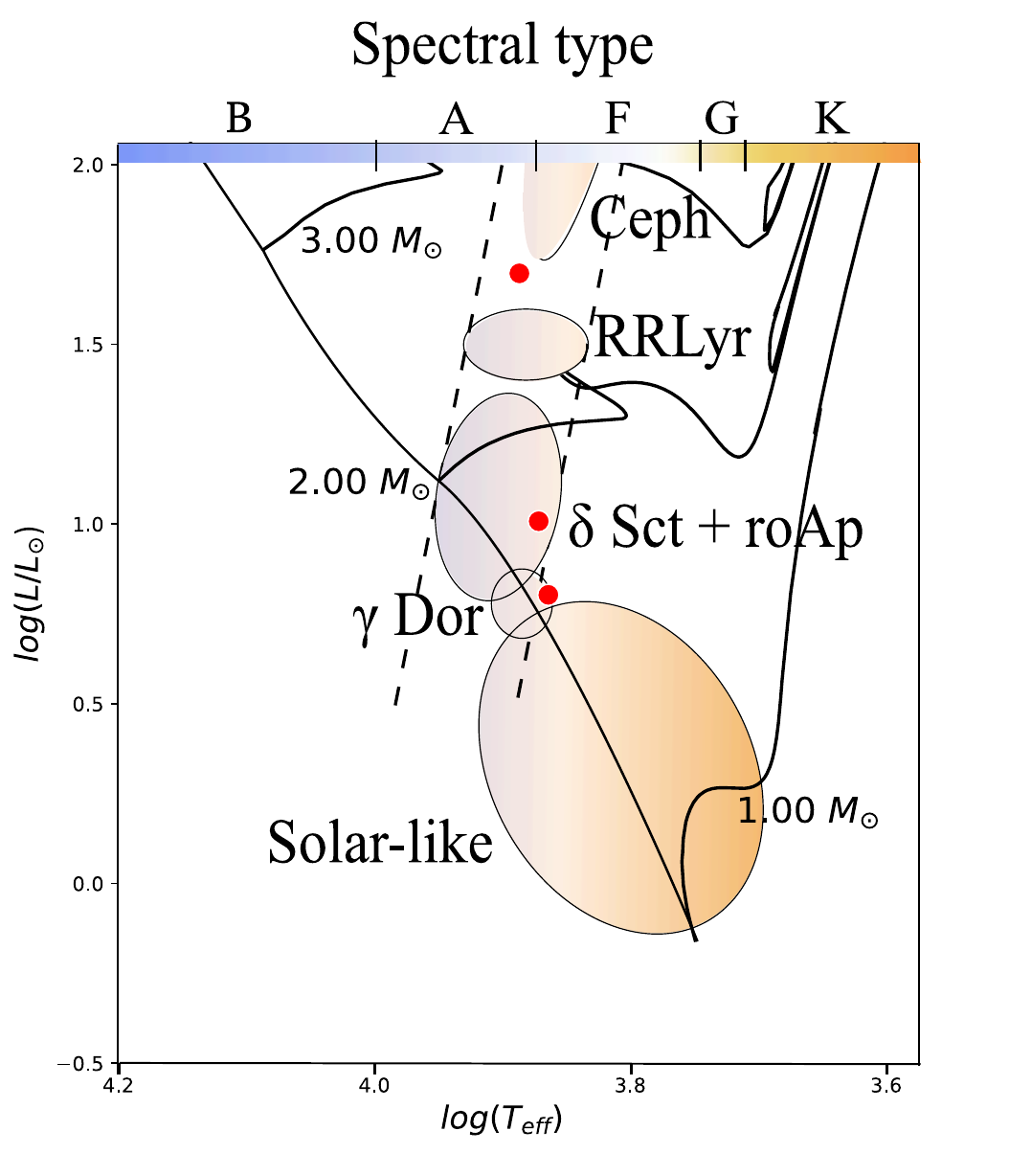}
	\caption{Hertzsprung-Russel (HR) diagram displaying the position of different classes of pulsating stars. Circles in red are for the three TOI pulsator candidates identified in the present study. Evolutionary model tracks are from \citet{Tang2014} (Figure adapted from \citealp{Aerts2021} and \citealp{Papics2013}).}
	\label{hrdiagram}
\end{figure*}

\section{Analysis of the pulsator candidates} \label{sec:results}

Based on a manifold procedure, which considers the analysis of TESS LCs taking into account Lomb-Scargle periodograms, FFT, and wavelet, we have identified three TOIs with likely pulsation signatures out of a sample of 3901 TOI stars. These TOI pulsator candidates are composed of one $\gamma$ Dor, TIC 164173105, and two $\delta$ Sct stars, TIC 118084044 and TIC 103195323. The location of the referred stars in the luminosity ($\log (L/L_{\odot})$) versus effective temperature ($T_{eff}$) diagram is displayed in Figure \ref{hrdiagram}, which also brings evolutionary tracks, at $[Fe/H] = 0.019$, from \citet{Tang2014} for masses ranging from 1.00 to 3.00$M_{\odot}$, as well as some loci of different classes of star pulsators adapted from \citet{Aerts2021} and  \citet{Papics2013}.

A few relevant steps are mandatories for the identification of stellar pulsating classes, namely the main pulsation frequency and its harmonics, the location in the HR Diagram, in comparison with the loci of known types of pulsators, and a comparison of the morphology of their LCs with well-known counterparts from the literature. In this context, we have computed the main pulsation frequency and its harmonics for each of the five stars considered bonafide pulsator candidates, including frequencies and linear combinations for multiple-mode stars, and searched for possible peaks of low amplitude and additional modes or modulation. The results of our analysis are given in Table 2, which lists the five main pulsation frequencies extracted for each star and eventual frequency linear combinations. The referred table also brings the probable pulsation class for each TOI star, based mainly on the comparison between the TOI LC, frequency range, and position in the HR Diagram, with known pulsators reported in the literature (\citealp{Cousins1992, Layden1997, Aerts2010, Pietrukowicz2013, Bravo2014, Bradley2015, Pakvstien2018, Paunzen2020}). In the following we discuss the different classes of pulsator candidates found in this study, including their contamination characteristics.

\subsection{Bonafide candidates}

\subsubsection{TIC 103195323}

This star was observed in sectors 25, 26, 52, 53, and 59. For high frequencies ($\approx$ 152-8671 $\mu$Hz) in all sectors, the RL $\approx$ 99\% indicates that the identified frequencies come from the same source (Gaia DR3 549998638456973184 or TIC 103195323). The periodicities 0.07 day (f $\approx$ 152 $\mu$Hz) and 0.032 day (f $\approx$ 357 $\mu$Hz) are in the range of the $\delta$ Sct Stars. The TPF analysis for each sector reinforces that the source of the frequencies is superimposed on the same target. The morphology of the LCs and its location in the HR Diagram are also compatible with the $\delta$ Sct pulsators.

\subsubsection{TIC 118084044}

This star was observed in sectors 35, 36, 37, 61, and 62. In sectors 35, 36, 37, and 61, the high frequencies ($\approx$142-8478 $\mu$Hz) come from the same source (Gaia DR3 5314833604298059136 or TIC 118084044) with RL = 99\%, which is confirmed by the TPF analysis. The periodicity of about 0.07 day (f $\approx$ 160 $\mu$Hz) and the morphology of the LCs are compatible with the $\delta$ Sct pulsators. Nevertheless, it is important to underline that TIC 118084044 presents contamination in sector 62 by the star Gaia DR3 5314833604301114240 with RL = 100.00\% for the referred high-frequency band. Otherwise, this star is a component of a resolved binary system, with a magnitude difference in the I band of 4.64 mag \citep{Ziegler2021}. Thus, the magnitude given in Table 1 and its position in the HR Diagram should be taken cautiously.

\subsubsection{TIC 164173105}

This star was observed in sectors 16 and 56. The analyzed frequencies ($\approx$ 1 and 40 $\mu$Hz) come from the same source (TIC 164173105 or Gaia DR3 1905671319879758464) with RL= 100.00\% for both sectors, which is reinforced by the TPF analysis. The periodicities 0.57 day or (f $\approx$ 19 $\mu$Hz), their LCs' morphology, and the HR Diagram's location are compatible with $\gamma$ Dor pulsators.

\subsection{Dubious or Contaminated Candidates}
\label{subdubious}

\subsubsection{TIC 58533991}

This star was only observed in sector 37. The \texttt{TESS\_localize} analysis shows that the target is not contaminated for high frequencies (around 120-8500 $\mu$Hz), with RL = 99.98\%. The periods of 0.085 day (f $\approx$ 121 $\mu$Hz) and 0.044 day (f $\approx$ 254 $\mu$Hz) are persistent in the wavelet maps and are located in the frequency range of the $\delta$ Sct stars. In addition, the morphology of the LC and the star's location in the HR Diagram are compatible with the $\delta$ Sct pulsators. A visual inspection of the TPF reinforces that these frequencies belong to the same source (Gaia DR3 6182008701312188672 or TIC 58533991). However, for low frequencies (approx. 3.5-32 $\mu$Hz), where the TOI orbital period (of 3.21 days) lies, there is potential contamination by the star Gaia DR3 6182008701312188800 (TIC 58533992). In addition, those periodic decays have depths compatible with a giant planet, but their morphology is compatible with an eclipsing binary.

\subsubsection{TIC 97700520}

This star was observed in sectors 33 and 34. Its frequency spectra and morphology of the LCs are compatible with an SPB variable type. In sector 33, the RL = 100.00\% corresponds to the same source (Gaia DR3 5604231696156773760 or TIC 97700520) for the low-frequency range (about 1.2-29 $\mu$Hz). However, the analysis of the TPF reveals that the source of origin of these frequencies is slightly displaced in relation to the considered target. In sector 34, this frequency range comes from two different sources: the star Gaia DR3 560423179235987200 (TIC 777293498) with RL = 65.30\% and the star Gaia DR3 56042310390220920 (TIC 97700515) with RL = 28.00\%.

\subsubsection{TIC 107782586}

This star was observed in sectors 34 and 61. Although this star shows the morphology of the LC, and periodicities from 0.4 to 1.4 days, compatible with an SPB variable type, the \texttt{TESS\_localize} analysis reveals contamination in both observed sectors. In sector 34, the contamination by the star Gaia DR3 5616803580833267968 (TIC 107782614) with RL = 86.40\% for low frequencies ($\approx$ 2-28 $\mu$Hz). In sector 61, there is contamination by the star Gaia DR3 5616803890070906624 (TIC 107782553) with RL = 74.96\%.

\subsubsection{TIC 123898871}

This star was observed only in sector 33. Its frequency spectra and the morphology of the LC are compatible with a $\gamma$ Dor variable type. Even though the low frequencies ($\approx$0.7-52 $\mu$Hz) come from the same source (Gaia DR3 2942380703200115200 or TIC 123898871) with RL = 100.00\%, the analysis of the TPF shows that the source of the referred frequencies is offset from the central target.

\subsubsection{TIC 149833117}

This star was observed in sectors 20, 47, and 60. Despite the observed periodicities and the morphology of the LCs to be compatible with a $\delta$ Sct variable, the analysis reveals a clear presence of contamination. For frequencies between 3 and 76 $\mu$Hz, the \texttt{TESS\_Localize} analysis shows evident contamination. In sector 20, there is contamination by the star Gaia DR3 1103137271667172224 (TIC 149833124), with RL = 100.00\%. In sector 47, the frequencies come from the source (Gaia DR3 1103137306027599744 or TIC 149833117) with RL = 20\%, and from another source (Gaia DR3 1103137306026909568 or TIC 743431 875), with RL = 45.84\%. In sector 60, the frequency signals come from TIC 743431875 with RL = 53.25\% and from the central source with RL = 46.75\%.

\subsubsection{TIC 150299840}

This star was observed in sectors 28, 29, 30, 31, 32, 33, 34, 35, 36, 38, 39, 61, and 62. Its frequency spectra and the morphology of the LCs are compatible with a $\gamma$ Dor variable type, but the \texttt{TESS\_Localize} analysis points to evident contamination. The frequencies extracted from the TESS LCs range from 1 to 31 $\mu$Hz. In sectors 28, 29, 30, 31, and 61, the \texttt{TESS\_Localize} analysis gives RL $\approx$ 99\% for contamination by the targets Gaia DR3 5481546402018694400 (TIC 150299832). In sectors 32 and 39, the contamination comes from the star Gaia DR3 5481546402018694272 (TIC 150300832), with RL = 100.00\%. In sector 33, the star has RL = 70.42\% from TIC 150299832 and RL = 29.58\% from TIC 150300832. In sector 34, the star has RL= 67.88\% from TIC 150299832 and RL = 32.12\% from TIC 150300832. In sector 35, the star has RL = 93.03\% from TIC 150299832 and RL = 6.97\% from TIC 150300832. In sector 36, the star has RL=83.55\% from TIC 150299832 and RL = 16.45\% from TIC 1503008 32. In sectors 38 and 62, the frequencies come from the source Gaia DR3 5481734177988936576 (TIC 150299840), with RL = 100.00\%, but the TPF analysis in these sectors shows that the frequency source is displaced from the central target.

\subsubsection{TIC 156987351}

This star was observed in sectors 6, 7, 33, 34, and 61. For all the sectors, the RL $\approx$ 100\%, and the source of the low frequencies ($\approx$15-90 $\mu$Hz) and the high frequencies ($\approx$ 115 at 16589 $\mu$Hz) come from the target (TIC 156987351 or Gaia DR3 5552450299121421696), a fact reinforced by the TPF analysis. The periodicities between 0.04 day (f $\approx$ 264 $\mu$Hz) and 0.26 day (f$\approx$ 41 $\mu$Hz), the morphology of the LCs and its location in the HR Diagram are compatible with the $\delta$ Sct pulsators.
Even though the pulsation signal potentially originates from the target, the TOI status is a false positive. According to \citet{Zhou2019}, the apparent transit signal in the LC of TIC 156987351 is likely a diluted signal of a faint eclipsing binary that is spatially blended with the target star in the photometric aperture.

\subsubsection{TIC 171160243}

This star was observed in sectors 43, 44, and 59. Its frequency spectra and the morphology of the LCs are compatible with a $\gamma$ Dor variable type, but evident contamination emerges from the \texttt{TESS\_Localize} analysis. For the low frequencies ($\approx$ 1-67 $\mu$Hz), the \texttt{TESS\_Localize} analysis reveals contamination by other sources in all the sectors. In sector 43, the contamination is due to the star Gaia DR3 172870238937744256 (TIC 171160247) with RL = 95.93\%. In sector 44, the contamination is due to the star Gaia DR3 172870204578002560 (TIC 171160242) with RL = 98.44\%. Finally, in sector 59, the contamination is due to the star Gaia DR3 172870170218265472 (TIC 661644392) with RL = 96.81\%. For high frequencies ($\approx$ 107-8086 $\mu$Hz), sectors 43 and 59 are also contaminated. In sector 43, the contamination comes from the star TIC 661644392 with RL= 80.7\%, and only RL = 19.29\% is due to the source of origin TIC 171160243 (Gaia DR3 172870170218265088). In sector 59, the analysis shows an RL = 85.28\% for the origin source, RL = 8.70\% for source TIC 661644392, and RL = 5.92\% for the Gaia DR3 172870273295899520 font (TIC 171160237). In sector 44, there is no contamination, with RL = 97.92\%. When analyzing the TPF in this sector, we observe that the origin of these frequencies is almost superimposed on the original target. However, the TPF analysis shows another GAIA target very close to the central star, making its pulsation classification ambiguous.

\subsubsection{TIC 179580045}

This star was observed in sectors 61 and 62. Its frequency spectra and the morphology of the LCs are compatible with a $\delta$ Sct variable type. However, the \texttt{TESS\_Localize} analysis shows that both sectors are contaminated in the extracted frequency range ($\approx$1 to 49 $\mu$Hz). In sector 61, the contamination comes from the star TIC 179580052 (Gaia DR3 4657936807635392512) with RL = 91.29\% and from star TIC 179580043 (Gaia DR3 4657936768966921984) with RL = 7.51\%. In sector 62, the contamination comes from three different sources: TIC 179579960 (Gaia DR3 4657936704556233344) with RL = 46.89\%, TIC 729054556 (Gaia DR3 4657936773275702272) with RL = 35.59\%, and TIC 179579957 (Gaia DR3 4657936395318580352) with RL = 17.52\%.

\subsubsection{TIC 281716779}

This star was observed only in sector 33. Its frequency spectra and the morphology of the LC are compatible with an RRLyrae variable type. Even though the \texttt{TESS\_Localize} analysis shows that the extracted frequencies ($\approx$1.8-32 $\mu$Hz) seem to come from the central source (Gaia DR3 3101977530394773504 or TIC 281716779), with RL = 99.99\%, the analysis of the TPF shows a slight offset of the source. We should be cautious with a possible pulsation classification.

\subsubsection{TIC 287196418}

This star was observed in sectors 14, 16, 17, 21, 26, 40, 41, 47, 49, 50, 51, 53, 54, 55, 56, 57, 58, 59, and 60. Its frequency spectra and the morphology of the LCs are compatible with a $\gamma$ Dor variable type. However, the \texttt{TESS\_Localize} analysis shows contamination of the source frequencies from $\approx$ 0.8 to 45 $\mu$Hz in the sectors 14, 16, 17, 26, 40, 41, 49, 51, 53, 54, 55, 56, 57, 58, and 60. In sector 14, the contamination comes from the star Gaia DR3 2238782528023282944 (TIC 287196425) with RL= 99.29\%. In sector 16, the contamination comes from the star Gaia DR3 2238782459303808256 (TIC 287196417) with RL= 99.66\%. In sector 17, the contamination comes from TIC 287196417 with RL = 87.03\%, while only 12.94\% probably comes from the source (TIC 287196418 or Gaia DR3 2238782528023283328). In sector 26, the contamination comes from TIC 287196417 with RL= 97.65\%, and only 2.35\% of the frequencies come from the source. In sector 40, the contamination comes from TIC 287196425 with RL=83.97\%) and TIC 287196423 (Gaia DR3 2238782528023282560) with RL= 16.01\%. In sector 41, 69.62\% of the frequencies come from the source and 13.60\% from TIC 287196419 (Gaia DR3 2238782455006216576). In sector 49, 46.86\% of the frequencies come from the source; 30.86\% originate from TIC 287196419, 19.95\% are from TIC 287196425, and 1.83\% are from TIC 287196417. In sector 51, the contamination originates from TIC 287196417, with RL = 98.16\%. In sector 53, the contamination comes from the star Gaia DR3 2238782562383024896 (TIC 287196405) with RL = 100.00\%. In sector 54, the contamination comes from the star TIC 287196425 with RL = 99.62\%. In sector 55, the contamination comes from TIC 287196425 with RL = 99.35\%. In sector 56, contamination comes from two sources: TIC 287196417 (RL = 71.73\%) and TIC 287196419 (RL = 28.05\%). In Sector 57, contamination comes from TIC 287196417 (RL = 55.54\%) and TIC 287196419 (RL = 44.46\%). In sector 58, the contamination comes from TIC 287196425, with RL = 95.04\%, and from TIC 287196423, with RL = 4.92\%. In sector 60, contamination comes from TIC 287196425, with RL = 96.58\%, and from TIC 287196419, with RL = 2.78\%.

\subsubsection{TIC 297967252}

This star was observed in sectors 9, 10, 35, 36, and 62. Its frequency spectra and the morphology of the LCs are compatible with an SPB variable type. However, the \texttt{TESS\_Localize} analysis shows that the source frequencies in the range 1-39 $\mu$Hz, are contaminated in all sectors. In sector 9, the contamination comes from the star TIC 297967275 (Gaia DR3 5312912791851514112) with RL = 100.00\%. In sector 10, the contamination comes from the star TIC 297967315 (Gaia DR3 5312912753187630592) with RL = 98.57\%. In sector 35, contamination comes from three different sources: TIC 297967226 (Gaia DR3 5312912894930645632) with RL = 80.85\%; TIC 297967211 (Gaia DR3 5312912894930645888) with RL = 14.18\%; TIC 297967259 (Gaia DR3 5312912890626593792) with RL = 4.09\%. In sector 36, the contamination comes from the star TIC 297967270 (Gaia DR3 5312912826211163648) with RL = 99.9\%. Finally, in sector 62, contamination comes from two different sources: TIC 859896818 (Gaia DR3 5312912894930644992) with RL = 64.58\% and TIC 297967231 (Gaia DR3 5312912894930644608) with RL = 31.84\%.

\subsubsection{TIC 333607525}

This star was observed only in sector 33. Its frequency spectra and the morphology of the LC are compatible with a $\gamma$ Dor variable type. However, the \texttt{TESS\_Localize} analysis shows that the extracted frequencies ($\approx$ 1-31 $\mu$Hz) are associated with contamination, where an RL = 50.59\% associated to the source itself (TIC 333607525 or Gaia DR3 2953400695928635648), but an RL= 47.20\% associated to another source (Gaia DR3 2953400695932187904).

\subsubsection{TIC 374095457}

The star TIC 374095457 was observed in sectors 9, 10, 36, and 37. Its frequency spectra and the morphology of the LCs are compatible with an RRLyrae/Cepheid variable type. However, the \texttt{TESS\_Localize} analysis shows that the identified frequencies ($\approx$ 0.85 to 59 $\mu$Hz) result clearly from contamination in all sectors. In sector 9, the frequencies appear to come from the central source (TIC 374095457 or Gaia DR3 5364393472448882304) with an RL = 72,47\%, but an RL = 27.53\% is associated with another source (Gaia DR3 5364393472439395328). In sector 10, the star is contaminated by this later source with RL= 91.77\% and RL = 8.23\% associated with the central target. In sector 36, there is a contamination in the frequency signal with an RL = 62.90\% associated with the central source and RL = 37.09\% associated with the source contaminating sectors 9 and 10. In sector 37, we find the RL = 63.57\% for the same source that contaminates the previous sectors and an RL= 36.43\% for the central source.

\subsubsection{TIC 436873727}

This star was observed in sectors 18, 42, 43, 44, and 58. Its frequency spectra and the morphology of the LCs are compatible with an RRLyr variable type. Even though the \texttt{TESS\_Localize} analysis shows that the analyzed frequencies ($\approx$ 1-68 $\mu$Hz) come from the central source (TIC 436873727 or Gaia DR3 114340658009875072), with an RL = 100.00\% for all the sectors (except sector 43), the TPF of these sectors shows that the frequency source is very far from the central target. Further, in sector 43, 94.8\% of the frequencies come from another source (Gaia DR3 114340829808567680 or TIC 436873726).

\subsubsection{TIC 468997317}

This star was observed in sectors 9, 10, 35, 36, and 62. Its frequency spectra and the morphology of the LCs are compatible with an SPB variable type, and the \texttt{TESS\_Localize} analysis shows that the frequency signals ($\approx$1-33 $\mu$Hz) are contaminated in all the sectors. In sector 9, 89.59\% of the signals come from the source (TIC 468997317 or Gaia DR3 5312673922948554880), and 10.41\% come from the source Gaia DR3 5312673922954351360. In sector 10, the signals come from this last source, with an RL = 58.89\%, and the other fraction comes from the central source, with an RL = 40.93\%. In sector 35, the contamination comes from the star Gaia DR3 5312673888594180608 (TIC 859817272) with RL = 81.65\% and from the star Gaia DR3 5312673922953919488 (TIC 468997323) with RL = 18.35\%. In sector 36, 68.76\% of the signals come from the central source, and 31.24\% come from another source (the same one that contaminates sectors 9 and 10). In sector 62, 80.29\% of the signals come from the source, and 19.67\% come from another source (the same one that contaminates sectors 9, 10, and 36).

\section{Summary and Conclusions} \label{sec:summary}

The high-cadence and the all-sky photometric survey of TESS Object of Interest is a valuable resource in the search for pulsation in planet host stars. We report the discovery of potential pulsation signatures in three TOIs that were monitored in a short cadence (2 minutes) by the TESS mission, all of them observed in more than one sector. The location of these stars in the HR Diagram, and the morphology of their LCs, associated with their pulsation frequencies, point to the following variable classes: two $\delta$ Sct (TIC 103195323 and TIC 118084044) and one $\gamma$ Dor (TIC 164173105). As far as we could find in the present-day literature, and considering the analysis of the available data, these stars have no indication of false-positive planet hosts. However, a spectroscopic follow-up is mandatory for a final confirmation on their classification as pulsating planet-host stars.

To date, a very scarce list of substellar companions orbiting pulsating stars is known. The diversity of pulsating stars hosting substellar companions presenting a variety of orbital periods, ranging from 2.74 to 5.59 days, will allow a search for traces of planet-induced pulsation. For instance, transient tidal effects can induce pulsation modes in the envelope of a star, but, to date, detecting such pulsations in planet-host stars is still very scarce \citep{deWit2017}. Furthermore, this diversity of pulsation variability, with multiple individual frequencies, will allow us to test the role of different stellar physical phenomena in pulsation, such as rotation, diffusion, and convection, in the presence of planet companions.

According to \citet{Sabotta2019}, there is little evidence for a high occurrence of close-in, massive planets around intermediate-mass stars. In this context, if the companions around the discovered TOI stars are confirmed as planets, then the present result will also contribute to our understanding of the presence of planets with short orbital period around intermediate-mass stars. Their locations in the HR diagram point to stars with mass between 1.3 and 2.0 solar masses and effective temperature corresponding to A-type stars. Indeed, the lack of close-in massive planets (with short periods) is today a crucial aspect in the discussion of planet formation theories \citep[e.g.,][]{Hasegawa2013,Stephan2018}.

The stable oscillations of pulsating stars can also serve as precise chronometers, which may be monitored for the effects of a planetary companion, produced by the gravitational tugs that they exert on the host star and other planets, causing periodic changes in pulsation arrival times (e.g., \citealp{Hermes2018}). The TOI pulsator candidates that are reported here may also represent a sensitive timing laboratory for detecting substellar companions, in particular around the hottest pulsating stars. The referred stars are also suitable for asteroseismic probing, which can further constrain the mass of the host stars and provide a more in-depth analysis of their atmosphere.

Finally, our study shows how challenging the analysis of TESS data is to avoid source confusion in the definition of the root cause of variability signature, a fact recently reported by \citet{Pedersen2023} and \citet{Mullally2022}. From a sample of 19 stars with apparent pulsator signatures in their LCs, we found consistent source locations for only three. For the other 16 targets, 10 present pulsating frequency signals contaminated by nearby sources, whereas four present offsets and two with contamination and offset effects.

\begin{acknowledgments}
        
We warmly thank our families for involving us with care, patience, and tenderness, during the home office tasks for the preparation of this study in the face of this COVID-19 difficult moment. Research activities of the observational astronomy board at the Federal University of Rio Grande do Norte are supported by continuous grants from the Brazilian funding agencies CNPq, FAPERN, and INCT-INEspa\c{c}o. This study was financed in part by the Coordena\c{c}\~ao de Aperfei\c{c}oamento de Pessoal de N\'ivel Superior - Brasil (CAPES) - Finance Code 001, and by Funda\c{c}\~ao de Amparo \`a Pesquisa do Estado de S\~ao Paulo (FAPESP) - grant 2016/13750-6. R.L.G. and Y.S.M. acknowledge CAPES graduate fellowships, and A.B.B. acknowledges CNPq graduate fellowship. L.A.A. thanks the CNPq for funding process (315502/2021-5). B.L.C.M., E.J.P., I.C.L., and J.R.M. acknowledge CNPq research fellowships. C.E.F.L. acknowledges the support provided by ANID’s Millennium Science Initiative through grant ICN12\_12009, awarded to the Millennium Institute of Astrophysics (MAS), and the support of ANID/FONDECYT Regular grant 1231637.  D.H. acknowledges support from ANID/doctoral fellowship grant 21232262. This paper includes data collected by the TESS mission. Funding for the TESS mission is provided by the NASA Explorer Program.

\end{acknowledgments}

\end{document}